# Picometer Registration of Zinc Impurity States in Bi$_2$Sr$_2$CaCu$_2$O$_{8+\delta}$ for Phase Determination in Intra-unit-cell Fourier Transform STM

M.H. Hamidian[1,2§], I. A. Firmo[1,2§], K. Fujita[1,2,3], S. Mukhopadhyay[1,2], J.W. Orenstein[4], H. Eisaki[5], S. Uchida[3], M.J. Lawler[2], E.-A. Kim[2] and J.C. Davis[1,2,6,7]

1. CMPMS Department, Brookhaven National Laboratory, Upton, NY 11973, USA.
2. Laboratory of Solid State Physics, Department of Physics, Cornell University, Ithaca, NY 14853, USA.
3. Department of Physics, University of Tokyo, Bunkyo-ku, Tokyo 113-0033, Japan.
4. Department of Physics, University of California, Berkeley, CA 94720, USA.
5. Institute of Advanced Industrial Science and Technology, Tsukuba, Ibaraki 305-8568, Japan.
6. School of Physics and Astronomy, University of St. Andrews, St. Andrews, Fife KY16 9SS, UK.
7. Kavli Institute at Cornell for Nanoscale Science, Cornell University, Ithaca, NY 14853, USA.
§   Contributed equally to this project.

**Direct visualization of electronic-structure symmetry within each crystalline unit cell is a new technique for complex electronic matter research[1-3]. By studying the Bragg peaks in Fourier transforms of electronic structure images, and particularly by resolving both the real and imaginary components of the Bragg amplitudes, distinct types of intra-unit cell symmetry breaking can be studied. However, establishing the precise symmetry point of each unit cell in real space is crucial in defining the phase for such Bragg-peak Fourier analysis. Exemplary of this challenge is the high temperature superconductor Bi$_2$Sr$_2$CaCu$_2$O$_{8+\delta}$ for which the surface Bi atom locations are observable, while it is the invisible Cu atoms that define the relevant CuO$_2$ unit-cell symmetry point. Here we demonstrate, by imaging with picometer precision the electronic impurity states at individual Zn atoms substituted at Cu sites, that the phase established using the Bi lattice produces a ~2%(2π) error relative to the actual Cu lattice. Such a phase assignment error would not diminish reliability in the determination of intra-unit-cell rotational symmetry breaking at the CuO$_2$ plane[1-3]. Moreover, this type of impurity atom substitution at the relevant symmetry site can be of general utility in phase determination for Bragg-peak Fourier analysis of intra-unit-cell symmetry.**

**CONTENTS**

**1      Spectroscopic Imaging STM and its Bragg-peak Fourier Analysis**

In spectroscopic imaging scanning tunneling microscopy (SI-STM), the differential tunneling conductance $dI/dV(\boldsymbol{r}, V) \equiv g(\boldsymbol{r}, V)$ between the tip and sample is measured as a function of both location $\boldsymbol{r}$ and electron energy $E=eV$. For a simple metal, $g(\boldsymbol{r}, V) = LDOS(\boldsymbol{r}, E = eV)$ where $LDOS(\boldsymbol{r}, E)$ is the spatially and energy resolved local-density-of-electronic-states [4]. This direct assignment cannot be made for materials whose electronic structure is strongly heterogeneous at the nanoscale [5,6]. However, even in those circumstances distances (wavelengths) and spatial symmetries in the $g(\boldsymbol{r}, V)$ images should retain their physical significance. SI-STM has proven to be of wide utility and growing significance in electronic structure studies, especially in situations where the translationally invariant non-interacting single-particle picture does not hold [6-14].

A key practical challenge for all such SI-STM studies is that, over the week(s) of continuous data acquisition required to measure a $g(\boldsymbol{r}, V)$ data-set having both ≥50 pixels within each crystal unit cell and the large field of view (FOV) required for precision Fourier analysis, thermal and mechanical drifts distort the $g(\boldsymbol{r}, V)$ images subtly. Such $g(\boldsymbol{r}, V)$ distortions, while pervasive, are usually imperceptible in conventional analyses. However, they strongly impact the capability to determine intra-unit-cell symmetry breaking because the perfect lattice-periodicity throughout $g(r,V)$ that is necessary for Bragg-peak Fourier analysis, is disrupted [1].

To address this issue, we recently introduced a post-measurement distortion correction technique that is closely related to an approach we developed earlier to address incommensurate crystal modulation effects [15]. The new procedure [1-3,16] identifies a slowly varying field $u(r)$ [17] that measures the displacement vector $u$ of each location $r$ in a topographic image of the crystal surface $T(r)$, from the location $r-u(r)$ where it should be if $T(r)$ were perfectly periodic with the symmetry and dimensions established by X-ray or other scattering studies of the same material.

To understand the procedure, consider an atomically resolved topograph $T(r)$ with tetragonal symmetry. In SI-STM, the $T(r)$ and its simultaneously measured $g(r,V)$ are specified by measurements on a square array of pixels with coordinates labeled $r$=(x,y). The power-spectral-density (PSD) Fourier transform of $T(r)$, $|\tilde{T}(q)|^2$ - where $\tilde{T}(q) = Re\ \tilde{T}(q) + iIm\ \tilde{T}(q)$, then exhibits two distinct peaks representing the atomic corrugations. These are centered at the first reciprocal unit cell Bragg wavevectors $Q_a = (Q_{ax}, Q_{ay})$ and $Q_b = (Q_{bx}, Q_{by})$ with $a$ and $b$ being the unit cell vectors. Next we apply a computational 'lock-in' technique in which $T(r)$ is multiplied by reference cosine and sine functions with periodicity set by the wavevectors $Q_a$ and $Q_b$, and whose origin is chosen at an apparent atomic location in $T(r)$. The resulting four images are filtered to retain only the $q$-space regions within a radius $\delta q = \frac{1}{\lambda}$ of the four Bragg peaks; the magnitude of $\lambda$ is chosen to capture only the relevant image distortions (in particular, $\delta q$ is chosen here to be smaller than the $Bi_2Sr_2CaCu_2O_{8+\delta}$ supermodulation wavevector). This procedure results in retaining the local phase information $\theta_a(r), \theta_b(r)$ that quantifies the local displacements from perfect periodicity:

$$X_a(r) = \cos\theta_a(r), Y_a(r) = \sin\theta_a(r) \quad (1a)$$
$$X_b(r) = \cos\theta_b(r), Y_b(r) = \sin\theta_b(r) \quad (1b)$$

Dividing the appropriate pairs of images then allows one to extract

$$\theta_a(r) = \tan^{-1}\frac{Y_a(r)}{X_a(r)} \quad (2a)$$

$$\theta_b(\mathbf{r}) = \tan^{-1}\frac{Y_b(\mathbf{r})}{X_b(\mathbf{r})} \tag{2b}$$

Of course, in a perfect lattice the $\theta_a(\mathbf{r}), \theta_b(\mathbf{r})$ would be independent of $\mathbf{r}$. However, in the real image $T(\mathbf{r})$, $\mathbf{u}(\mathbf{r})$ represents the distortion of the local maxima away from their expected perfectly periodic locations, with the identical distortion occurring in the simultaneous spectroscopic data $g(\mathbf{r}, V)$. Considering only the components periodic with the lattice, the measured topograph can therefore be represented by

$$T(\mathbf{r}) = T_0[\cos(\mathbf{Q}_a \cdot (\mathbf{r}+\mathbf{u}(\mathbf{r}))) + \cos(\mathbf{Q}_b \cdot (\mathbf{r}+\mathbf{u}(\mathbf{r})))] \tag{3}$$

Correcting this for the spatially dependent phases $\theta_a(\mathbf{r}), \theta_b(\mathbf{r})$ generated by $\mathbf{u}(\mathbf{r})$ requires an affine transformation at each point in (x,y) space. From Eqn. 3 we see that the actual local phase of each of cosine component at a given spatial point $\mathbf{r}$, $\varphi_a(\mathbf{r}), \varphi_b(\mathbf{r})$, can be written as

$$\varphi_a(\mathbf{r}) = \mathbf{Q}_a \cdot \mathbf{r} + \theta_a(\mathbf{r}) \tag{4a}$$
$$\varphi_b(\mathbf{r}) = \mathbf{Q}_b \cdot \mathbf{r} + \theta_b(\mathbf{r}) \tag{4b}$$

where $\theta_i(\mathbf{r}) = \mathbf{Q}_i \cdot \mathbf{u}(\mathbf{r}); i = a, b$ is the additional phase generated by the distortion field $\mathbf{u}(\mathbf{r})$. This simplifies Eqn. 2 to

$$T(\mathbf{r}) = T_0[\cos(\varphi_a(\mathbf{r})) + \cos(\varphi_b(\mathbf{r}))] \tag{5}$$

which is defined in terms of its local phase fields only, and every peak associated with an atomic local maximum in the topographic image has the same $\varphi_a$ and $\varphi_b$. The goal is then to find a transformation, using the given phase information $\varphi_{a,b}(\mathbf{r})$, to map the distorted lattice onto a perfectly periodic one. This is equivalent to finding a set of local transformations which makes $\theta_{a,b}$ take on constant values, $\bar{\theta}_a$ and $\bar{\theta}_b$, over all space. Thus, let $\mathbf{r}$ be a point on the unprocessed (distorted) $T(\mathbf{r})$, and let $\tilde{\mathbf{r}} = \mathbf{r} - \mathbf{u}(\mathbf{r})$ be the point of equal phase on the 'perfectly' lattice-periodic image which needs to be determined. This produces a set of equivalency relations

$$\mathbf{Q}_a \cdot \mathbf{r} + \theta_a(\mathbf{r}) = \mathbf{Q}_a \cdot \tilde{\mathbf{r}} + \bar{\theta}_a$$
$$\mathbf{Q}_b \cdot \mathbf{r} + \theta_b(\mathbf{r}) = \mathbf{Q}_b \cdot \tilde{\mathbf{r}} + \bar{\theta}_b \tag{6}$$

Solving for the components of $\tilde{\mathbf{r}}$ and then re-assigning the $T(\mathbf{r})$ values measured at $\mathbf{r}$, to the new location $\tilde{\mathbf{r}}$ in the (x,y) coordinates produces a topograph with 'perfect' lattice periodicity. To solve for $\tilde{\mathbf{r}}$ we rewrite Eqn. 6 in matrix form:

$$Q\begin{pmatrix}\tilde{r}_1\\\tilde{r}_2\end{pmatrix} = Q\begin{pmatrix}r_1\\r_2\end{pmatrix} - \begin{pmatrix}\bar{\theta}_a - \theta_a(r)\\\bar{\theta}_b - \theta_b(r)\end{pmatrix} \tag{7}$$

where
$$Q = \begin{pmatrix}Q_{ax} & Q_{ay}\\Q_{bx} & Q_{by}\end{pmatrix} \tag{8}$$

Because $Q_a$ and $Q_b$ are orthogonal, $Q$ is invertible allowing one to solve for the displacement field $u(r)$ which maps $r$ to $\tilde{r}$ as

$$u(r) = Q^{-1}\begin{pmatrix}\bar{\theta}_a - \theta_a(r)\\\bar{\theta}_b - \theta_b(r)\end{pmatrix} \tag{9}$$

In practice, we use the convention $\bar{\theta}_i = 0$ which generates a 'perfect' lattice with an atomic peak at the origin; this is equivalent to ensuring that there are no imaginary (sine) components to the Bragg peaks in the Fourier transform.

Using this technique, one can estimate *u(r)* and thereby undo distortions in the raw *T(r)* data with the result that it is transformed into a distortion-corrected topograph *T'(r)* exhibiting the known periodicity and symmetry of the termination layer of the crystal. The key step for electronic-structure symmetry determination is then that the identical geometrical transformations to undo *u(r)* in *T(r)* yielding *T'(r)*, are also carried out on every $g(r,V)$ acquired simultaneously with the *T(r)* to yield a distortion corrected $g'(r,V)$. The *T'(r)* and $g'(r,V)$ are then registered to each other and to the lattice with excellent periodicity. This procedure can be used quite generally with SI-STM data that exhibits appropriately high resolution in both *r*-space and *q*-space.

## 2  'Real' and 'Imaginary' Contributions to the Bragg Peaks in $\tilde{g}(q,V)$

Bragg-peak Fourier analysis of an electronic structure image $g(r,V)$ focuses upon $\tilde{g}(q,V) = Re\,\tilde{g}(q,V) + iIm\,\tilde{g}(q,V)$, its complex valued two-dimensional Fourier transform. Here $Re\,\tilde{g}(q,V)$ is the cosine and $Im\,\tilde{g}(q,V)$ the sine Fourier-component, respectively. By focusing on the Bragg peaks $q=Q_a,Q_b$ only those electronic phenomena with the same spatial periodicity as the crystal are considered. Obviously, successful application of this approach when using $g(r,V)$ images requires: (i) a highly accurate registry of the unit-cell origin to satisfy the extreme sensitivity in $\tilde{g}(q,V)$ to the phase, (ii) that the $g(r,V)$ data-set has adequate sub-unit-cell resolution without which distinctions between the four inequivalent Bragg amplitudes at $Q_a,Q_b$ would be zero and, (iii) that this same $g(r,V)$ be measured in a large FOV so as to achieve high resolution in $q$-space. Only recently has this combination of characteristics in $g(r,V)$ measurement been achieved [1-3].

With the availability of such data, several measures of intra-unit-cell breaking of crystal symmetry by the electronic structure become possible from the study of the real and imaginary components of the Bragg amplitudes in $\tilde{g}(Q,V)$. For example, if the crystal unit-cell is tetragonal with 90°-rotational ($C_{4v}$) symmetry, one can search for intra-unit-cell "nematicity" ( breaking of $C_{4v}$ down to 180°-rotational ($C_{2v}$) symmetry ) in the electronic structure by considering $O_N(V) = Re\ \tilde{g}(Q_b,V) - Re\ \tilde{g}(Q_a,V)$ [1-3]. Similarly, if the crystal unit-cell is centrosymmetric, one can search for intra-unit-cell breaking of inversion symmetry in electronic structure using $O_I(V) = |Im\ \tilde{g}(Q_a,V)| + |Im\ \tilde{g}(Q_b,V)|$. Obviously, however, in both of these cases and in general, the correct determination of Re $\tilde{g}(q,V)$ and Im $\tilde{g}(q,V)$ is critical. The assignment of the zero of coordinates at the symmetry point of the unit-cell (and thus the correct choice of phase) is therefore the fundamental practical challenge of Bragg-peak Fourier transform SI-STM.

## 3    Intra-unit cell electronic symmetry breaking in $Bi_2Sr_2CaCu_2O_{8+\delta}$

An excellent example of this challenge can be found in the copper-oxide high temperature superconductor $Bi_2Sr_2CaCu_2O_{8+\delta}$. In general, copper-oxide superconductors are 'charge-transfer' Mott insulators and are strongly antiferromagnetic due to inter-copper superexchange [18]. Doping these materials with a hole-density $p$ to create superconductivity is achieved by removing electrons from the O atoms in the $CuO_2$ plane [19,20]. Antiferromagnetism exists for $p < 2\text{-}5\%$, superconductivity occurs in the range 5-10% $< p <$ 25-30%, and a Fermi liquid state appears for $p >$ 25-30%. For $p<20\%$ an unusual electronic excitation with energy scale $|E|=\Delta_1$, and which is anisotropic in $k$-space [21-25], appears at temperature far above the superconducting critical temperature. This region of the phase diagram has been labeled the 'pseudogap' (PG) phase because the energy scale $\Delta_1$ could be the energy gap of a distinct electronic phase [21, 22].

Intra-unit-cell spatial symmetries of the E~$\Delta_1$ (PG) states can be imaged directly using SI-STM in underdoped cuprates [1-3,6,16]. Typically, the function $Z(\bm{r},V) = g(\bm{r},+V)/g(\bm{r},-V)$ is used because it eliminates the severe systematic errors in $g(\bm{r},V)$ generated by the intense electronic heterogeneity effects specific to these materials [1,2,3,5,6,11,16]. These $Z(\bm{r},E)$ images reveal compelling evidence for intra-unit-cell $C_{4v}$ symmetry breaking specific to the states at the E~$\Delta_1$ pseudogap energy [6]. However, for Bragg peak Fourier transform studies of this effect the choice of origin of the $CuO_2$ unit cell (and thus the phase of the Fourier transforms) was determined by using the imaged locations of the Bi atoms in the BiO layer [1-3], while it is knowledge of the actual Cu atom locations which is required to most confidently examine intra-unit-cell symmetry breaking of the $CuO_2$ unit cell.

## 4    Cu-lattice phase-resolution challenge in $Bi_2Sr_2CaCu_2O_{8+\delta}$

To identify these sites and thus the correct phase, we studied lightly Zn-doped $Bi_2Sr_2CaCu_2O_{8+\delta}$ crystals with $p$~20%. Each $g(\bm{r},E=eV)$ map required ~5 days and a typical resulting topograph *T(r)* with 64 pixels covering the area of each $CuO_2$ unit cell is shown in Fig. 1a. This is an unprocessed topographic image *T(r)* of the BiO layer with the bright dots occurring at the location of Bi atoms. The inset shows a tightly focused measurement at the location of one of the Bragg peaks in $|\tilde{T}(\bm{q})|^2$; this clearly has spectral weight distributed over numerous pixels indicating the imperfect nature of the periodicity in this *T(r)*. Figure 1b is the simultaneously measured image of electronic structure $g(\bm{r},E)$ determined near E~$\Delta_1$. Figure 1c shows the PSD Fourier transform of Figure 1b while its inset focuses upon a single Bragg peak. Figure 1d shows the processed topographic image *T'(r)* after distortion correction using Eqn. 9. The subtlety of these corrections is such that Fig. 1d appears virtually identical to Fig. 1a at first sight. However, the inset shows that the Bragg peak of the PSD Fourier transform $|\tilde{T}(\bm{q})|^2$ of Fig. 1d now becomes isotropic and consists of just a single pixel; this indicates that Bi atom periodicity is now as perfect as possible given the limitations of $q$-space resolution from the finite FOV. Figure 1e is the $g(r,V)$ data of Fig. 1b but now processed in the same distortion correction fashion as Fig. 1d to yield a function $g'(r,V)$. Its PSD Fourier transform $|\tilde{g}'(q,V)|^2$ as shown in Fig. 1f reveals how the Bragg peaks are now also sharp, indicating that the same spatial periodicity now exists in the electronic structure images (inset Fig. 1e). Nevertheless, the location of the Cu sites in the CuO$_2$ plane cannot be determined from the BiO *T'(r)* and therefore the phase for Bragg-peak Fourier analysis of $\tilde{g}'(q,V)$ from the CuO2 plane retains some ambiguity.

## 5     Imaging the electronic impurity state at Zn atoms substituted for Cu

To directly identify the symmetry point of the CuO$_2$ unit cell in a BiO topograph, we image $g(r, V = -1.5\text{mV})$ measured on Zn-doped Bi$_2$Sr$_2$CaCu$_2$O$_{8+\delta}$ crystals. A conductance map in a 60nm square region (simultaneous topograph Fig. 2a) is shown in Fig. 2b; the overall light background is indicative of a very low conductance near $E_F$ as expected in the superconducting state. However, we also detect a significant number of randomly distributed dark sites corresponding to areas of high conductance each with a distinct four-fold symmetric shape and the same relative orientation. The spectrum at the center of a dark site has a very strong intra-gap conductance peak at energy E=-1.5±0.5 meV, while the superconducting coherence peaks are suppressed [26]. This is a unitary strength quasiparticle scattering resonance at a single, potential-scattering, impurity atom in a *d*-wave superconductor [26,27]. These signatures can be used to identify the location of Zn atoms substituted on the Cu sites of Bi$_2$Sr$_2$CaCu$_2$O$_{8+\delta}$.

Figure 2a actually shows the topographic image *T'(r)* of the BiO layer *after* its distortion correction has been carried out, while Figure 2b shows the identically distortion corrected image of differential conductance. Fourteen Zn impurity state sites at Cu sites in the CuO$_2$ plane are observed. Imaging the locations of these individual Zn resonance sites with ~picometer resolution allows precision identification of the symmetry point of each CuO$_2$ unit cell and an excellent estimate of the correct phase required for Bragg-peak Fourier analysis of $CuO_2$ plane electronic structure.

## 6  Determination of Cu-lattice phase error from Bi-lattice calibration

In Figure 3, each pair of panels a-b, c-d, …, w-x, contains the simultaneously measured and identically distortion-corrected images $T''(r)$ and $g'(r, -1.5mV)$, each with 76 pixels inside the area of every $CuO_2$ unit cell. The coordinates of every Bi atom in the perfectly square lattice are known with ~pm precision in these $T''(r)$ images. The location of the Zn impurity state in each of the $g'(r, -1.5mV)$ images is determined by fitting a two-dimensional Gaussian to the central peak of the Zn resonance; a typical resulting error-bar for the location of the maximum is a value between 1 and 2 pm (SI Section I). The smallness of this error with respect to the pixel size is well understood in terms of the large signal to noise ratio at the Zn resonances [26] (SI section I). These procedures yield the displacement vector $d$ of every Zn-resonance maximum from the site of the nearest Bi atom as identified in $T''(r)$.

Figure 4a indicates using a black arrow the displacement vector $d$ between the Bi atom (top) and Cu atom (bottom) if a rigid xy-displacement existed between them. The blue planes indicate the relative positions of the two layers if there were no such shift and thus $d$=0. Figure 4b shows combined analysis of the measured values of $d$ for all the $CuO_2$ unit cells containing a Zn atom (with the origin of each centered at the relevant Bi atom identified from the nearest maximum in $T''(r)$ images in Figure 3).  The measured $d$ of every Zn resonance is shown as a red dot. The resulting average (Zn, Bi) displacement vector shown in black has a magnitude of ~2% of the $CuO_2$ unit cell dimension (1 standard deviation of the distribution is indicated by the grey ellipse). It is quite obvious from Fig. 4b that the Zn resonances are extremely close to the Bi sites, meaning that the $CuO_2$ layer is not shifted significantly from its expected location below the BiO layer (Fig. 4a).

Beyond the fact that the average displacement <$d$> represents only a ~2%(2π) error in the phase determination for the $CuO_2$ layer when using the BiO layer, other information on systematic errors within the SI-STM approach can be examined using these data. For example, the observed distribution of **d** rules out the existence of any discrete pixel displacement between *T(r)* and its simultaneously measured $g(r,V)$, as might occur if there were a software or processing error. Another point is that any spatial drift of the tip location during the hundreds of elapsed milliseconds between the measurement of the topographic signal and the differential conductance signal is also below 2% of the unit cell dimension. In fact the data in Fig.s 2,3,4 show that, for our instruments the measured *T(r)* and $g(r, V = -1.5 mV)$ are registered to each other within a few pm. Additionally, studies of this same set of Zn resonances using the 180º rotated scan direction (SI Section II) yield an equivalently narrow (but distinct) distribution of values of **d**. Moreover the center of this distribution is not shifted along the scan relative to that in Fig. 4b, indicating that random picometer scale image distortions dominate the **d** distribution and not the trajectory of the tip. Thus we do not currently regard the apparent displacement (CuO$_2$,BiO) in Fig. 4b as a property of the crystal lattice, but rather due to measurement limitations at these picometer length scales (see SI Section II).

## 7 Conclusions and Future

Three key practical conclusions emerge from these studies. First, the lateral shift between the surface BiO layer and the CuO$_2$ layer is measured at less than 2% of the unit cell dimension. Second, at this ~picometer precision there is no resolvable spatial drift of the tip location during the fractions of a second between the measurement of the topographic signal and the differential conductance signal. Perhaps most importantly, the ~2%(2π) phase error in the choice of the Cu-lattice origin observed here would not, based on results from our simulations (SI Section III), impact Fourier transform analysis using $Re\ \tilde{g}(Q_a, V)$ and $Re\ \tilde{g}(Q_b, V)$ to determine symmetry breaking in $g(r, V)$. However, such a ~2%(2π) phase error would produce a significant effect on a measure of intra-unit-cell inversion-symmetry breaking like $O_I(V)$, yielding an incorrect non-zero value for $Im\ \tilde{g}(Q_a, V)$ or $Im\ \tilde{g}(Q_b, V)$ of ~15% of $Re\ \tilde{g}(Q_a, V)$ or $Re\ \tilde{g}(Q_b, V)$ (SI Section III). Specifically, a ~2%(2π) phase assignment error for the Cu sites does not diminish reliability in the determination of intra-unit-cell rotational symmetry breaking at the $CuO_2$ plane[1-3]. Of more long term significance is that impurity atom substitution at the relevant symmetry site as demonstrated here, can be of general utility in accurate phase determination for Bragg-peak Fourier analysis of intra-unit-cell symmetry.

**Acknowledgements**

We are particularly grateful for help and advice from J.E. Hoffman. We acknowledge and thank A.V. Balatsky, D.-H. Lee, Kyungmin Lee, K. McElroy, S. Sachdev, J. Sethna, A. Schmidt and J. Zaanen for discussions and communications. These studies were supported by the U.S. Department of Energy, Office of Basic Energy Sciences. HE and SU acknowledge support from Grant-in-Aid for Scientific Research from the Ministry of Science and Education (Japan) and the Global Centers of Excellence Program for Japan Society for the Promotion of Science. IAF acknowledges support from Fundação para a Ciência e a Tecnologia, Portugal under fellowship number SFRH/BD/60952/2009.

**Figure Captions**

**Figure 1**

a)  Raw topographic image $T(r)$ of the BiO layer of $Bi_2Sr_2CaCu_2O_{8+\delta}$ in a 20nm FOV. The inset shows one of the Bragg peaks. All the data in this Figure were acquired at 1 G$\Omega$ junction resistance, -110 mV tip-sample bias.

b)  Unprocessed differential conductance image $g(r, E/\Delta_1(r) = 1)$ at the superconducting coherence peak energy. This image has been measured in the same FOV as a).

c)  PSD Fourier transform of b), $|\tilde{g}(q, e = 1)|^2$. Inset shows the same Bragg peak as a). The red circle indicates the location of the Bragg peak shown in the insets of Figure 1a),c),d),f).

d)  Same topographic image $T''(r)$ as in a), but after applying the distortion correction algorithm. The inset shows one of the resulting isotropic Bragg peaks confined in a single pixel (same as in a).

e)  Same differential conductance data $g'(r, e = 1)$ as in part b after applying the distortion correction algorithm.

f)  PSD Fourier transform of part e), $|\tilde{g}'(q, e = 1)|^2$. The red circle indicates the location of the Bragg peak shown in the insets of Figure 1a),c),d),f).

**Figure 2**

a) A ~60 nm square FOV *T''(r)*. The data shown in this image have been processed using the distortion correction algorithm.

g) Simultaneously measured $g'(r, V = -1.5\text{mV})$ in the same FOV as part a. Fourteen Zn impurity resonances are distinguishable in this image. These data have been processed using the distortion correction algorithm. The data in a and b were acquired at 1 GΩ junction resistance, 60 mV tip-sample bias.

b)

**Figure 3** a) - b) Simultaneous *T''(r)* image of a 60 Å FOV containing one Zn atom. The red cross indicates the Bi atom nearest to the Zn atom (in b); and $g'(r, eV)$ - this image was measured in the same FOV as a. The data in this Figure were acquired at 1 GΩ junction resistance, 60 mV, and was obtained from a double layer $g(r, E)$ map. A total time of 340ms has elapsed between the measurement of a) and b). All subsequent image pairs represent the equivalent data at a different location. All data in this Figure were obtained from five maps with identical acquisition parameters, and have been processed using the distortion correction algorithm.

**Figure 4.**

a) Schematic relationship between Bi atom locations in *T''(r)* and the Cu sites at the symmetry point of the CuO$_2$ until cell. The Bi (blue) atoms are directly above the Cu (green) atoms; in between the two atoms, there is an oxygen atom (not pictured here). Sometimes, a Zn impurity atom (red) is found at one of the Cu sites. Vector **d** is non-zero when there is small shift in the XY-plane between the positions of a Cu (or Zn impurity) atom and the Bi atom directly above it.

b) Displacement vector between center point of each Zn resonance in $g'(r, eV)$ and the nearest Bi atom in *T''(r)* (see Figure 3). The average displacement vector over all the (Zn, Bi) XY-plane distances is shown as a black arrow. The ellipse about the average displacement arrow corresponds to one standard deviation of the average displacement. The individual (Zn, Bi) XY-plane displacement data are shown by red dots. The fast and slow scan directions are respectively indicated by the thicker and thinner arrows on the bottom right corner of this Figure.

Figure 1

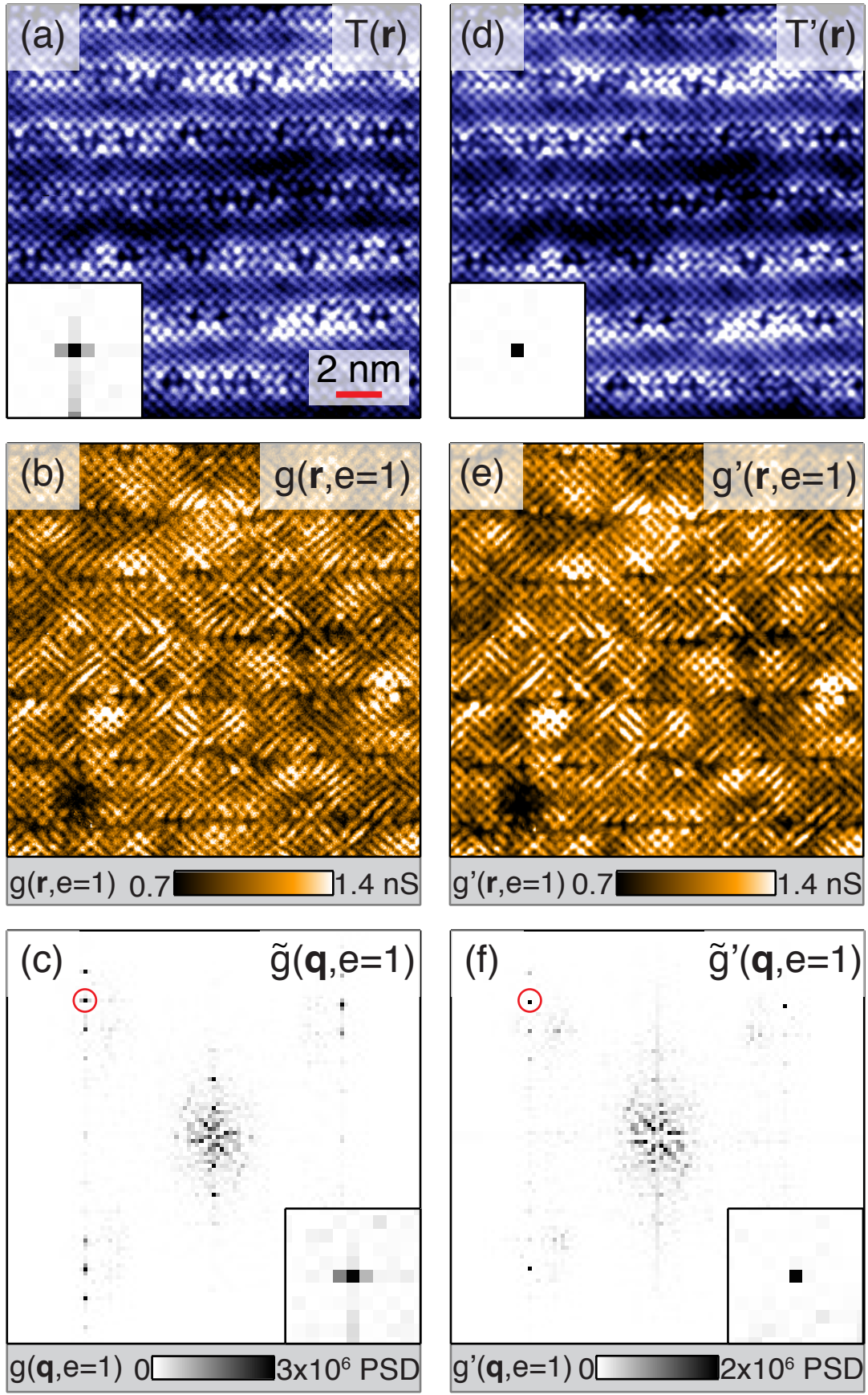

Figure 2

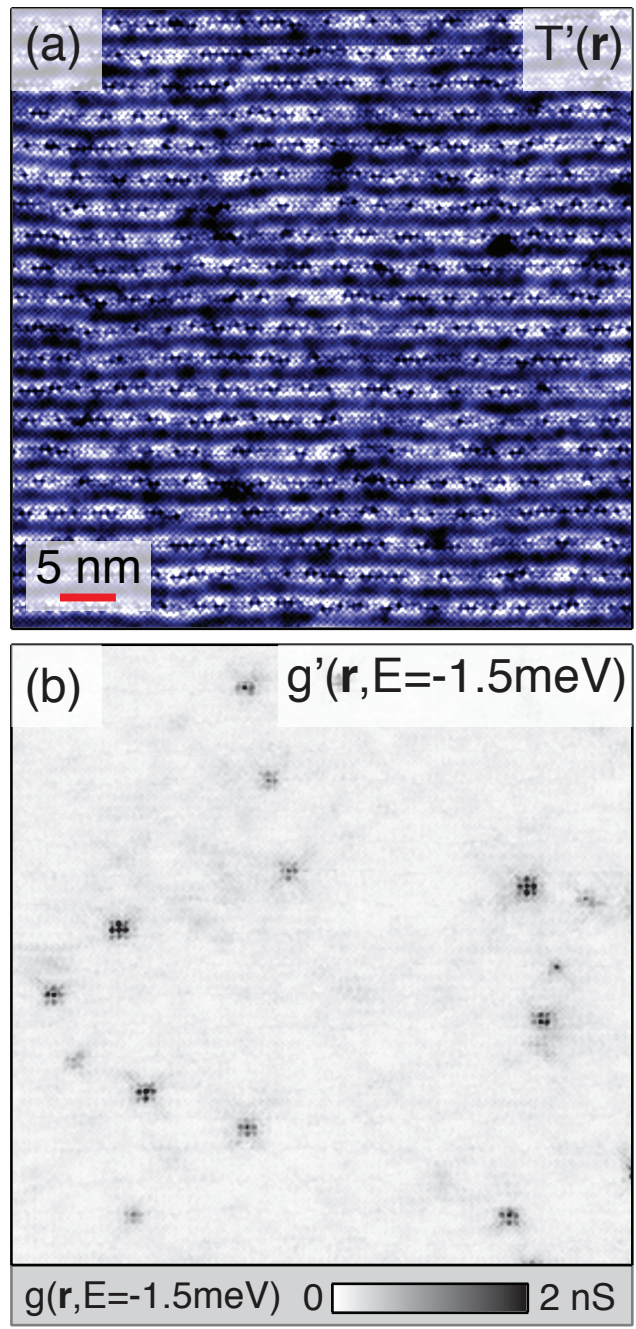

Figure 3

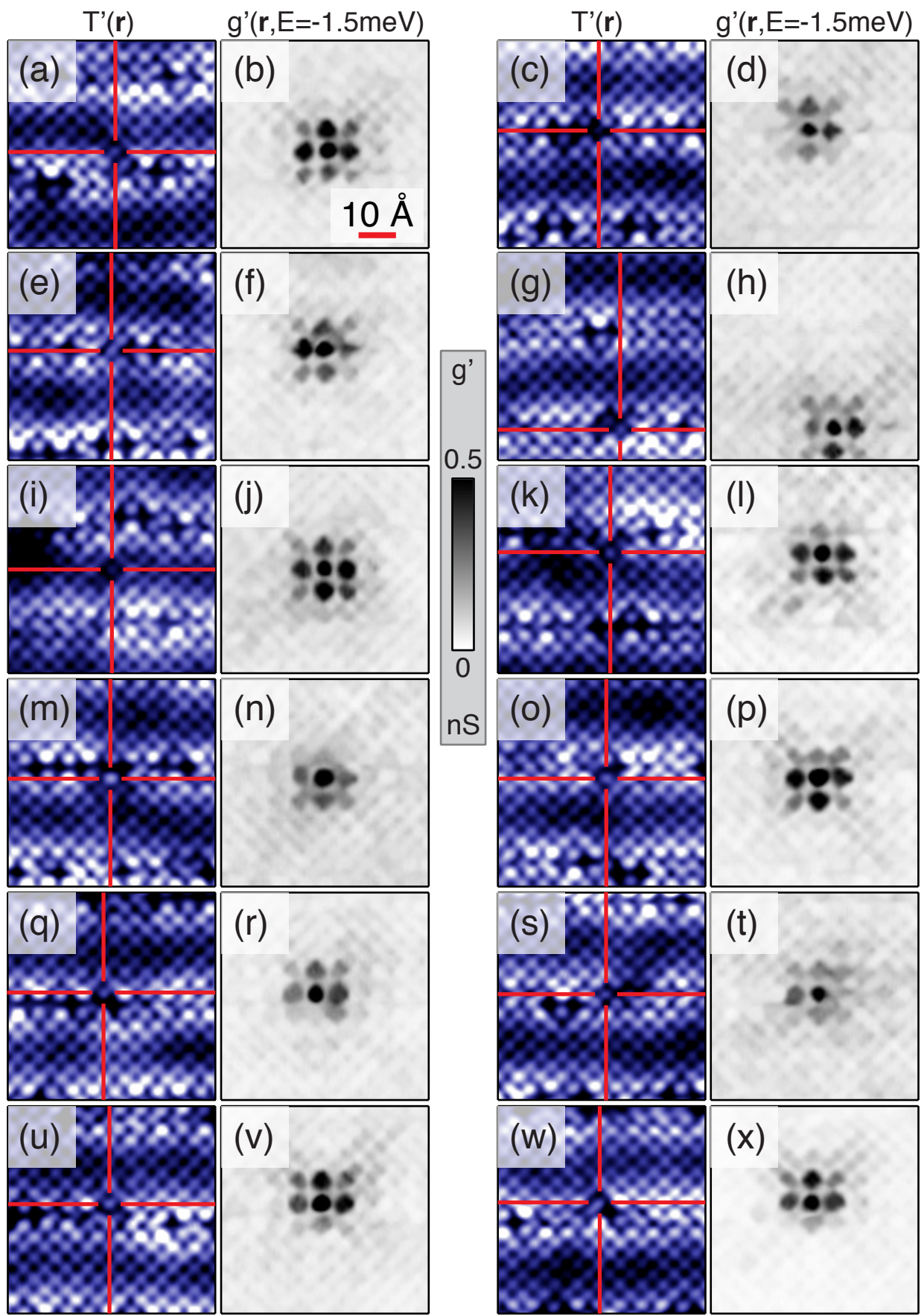

Figure 4

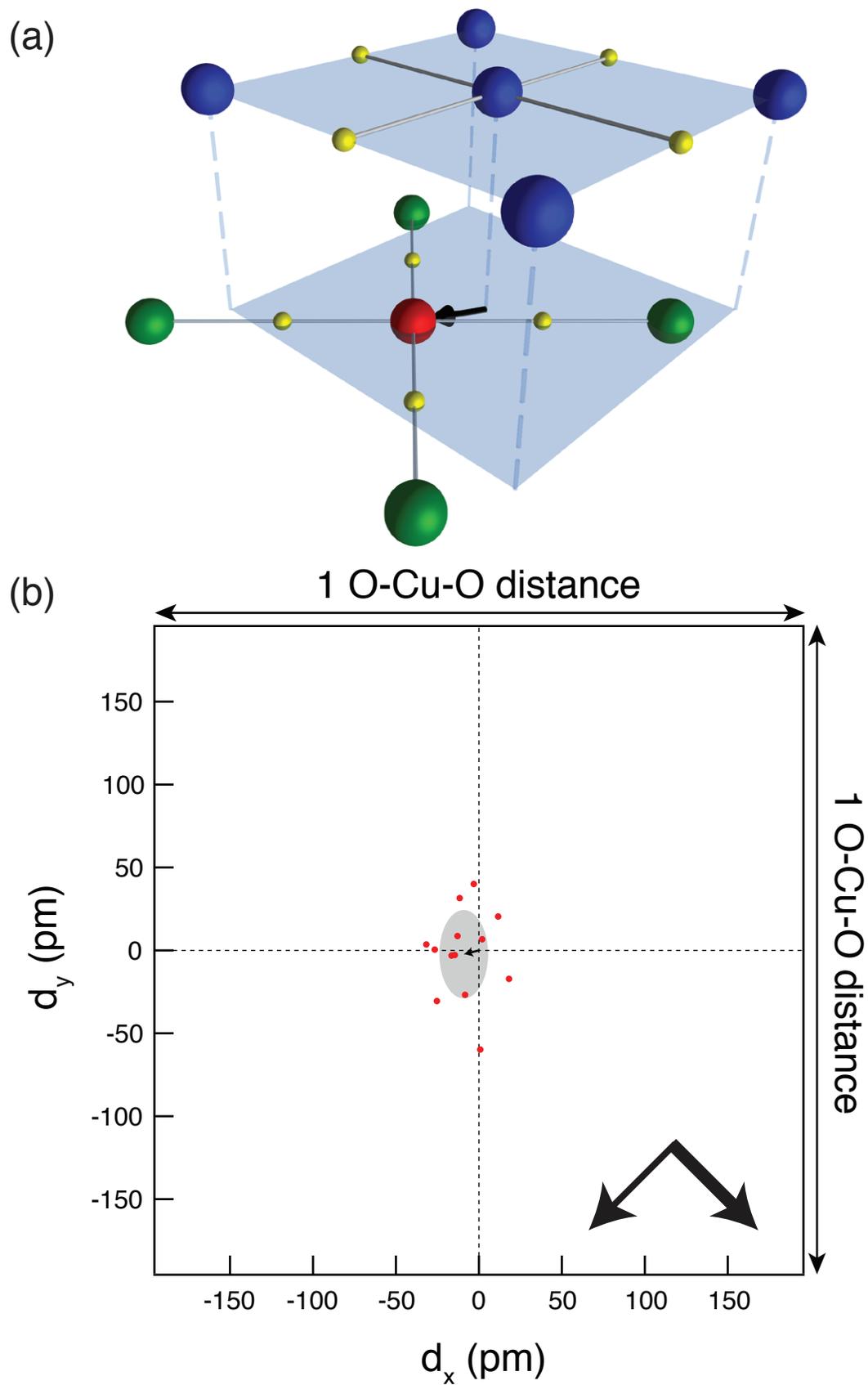

Supplementary Data for

# Picometer Registration of Zinc Impurity States in $Bi_2Sr_2CaCu_2O_{8+\delta}$ for Phase Determination in Intra-unit-cell Fourier Transform STM

M.H. Hamidian[1,2§], I. A. Firmo[1,2§], K. Fujita[1,2,3], S. Mukhopadhyay[1,2], J.W. Orenstein[4], H. Eisaki[5], S. Uchida[3], M.J. Lawler[2], E.-A. Kim[2] and J.C. Davis[1,2,6,7]

1. CMPMS Department, Brookhaven National Laboratory, Upton, NY 11973, USA.
2. Laboratory of Solid State Physics, Department of Physics, Cornell University, Ithaca, NY 14853, USA.
3. Department of Physics, University of Tokyo, Bunkyo-ku, Tokyo 113-0033, Japan.
4. Department of Physics, University of California, Berkeley, CA 94720, USA.
5. Institute of Advanced Industrial Science and Technology, Tsukuba, Ibaraki 305-8568, Japan.
6. School of Physics and Astronomy, University of St. Andrews, St. Andrews, Fife KY16 9SS, UK.
7. Kavli Institute at Cornell for Nanoscale Science, Cornell University, Ithaca, NY 14853, USA.
§ Contributed equally to this project.

### I. Quality of Gaussian fits to Zn resonances

The location of every Zn atom was obtained by fitting a two-dimensional (2D) Gaussian function to the center hub of each Zn resonance, where the 2D Gaussian was not constrained to be isotropic. Each pixel has a lateral dimension of 44pm, but a typical Gaussian fit error is ~1pm. The very high signal to noise ratio at the Zn resonances [1] accounts for the order of magnitude improvement in spatial refinement, because of the high quality of the fits. Figure S1 shows representative line cuts along the center resonance of a Zn atom as black squares. To illustrate the fitting scheme, also shown are 1D Gaussian fits to the data (red). The agreement between fits and data is clear. For other Zn resonances reported in Figure 3, both the signal to noise in the central resonance and the 1D or 2D Gaussian fits are of equivalent quality as the one reported in Figure S1.

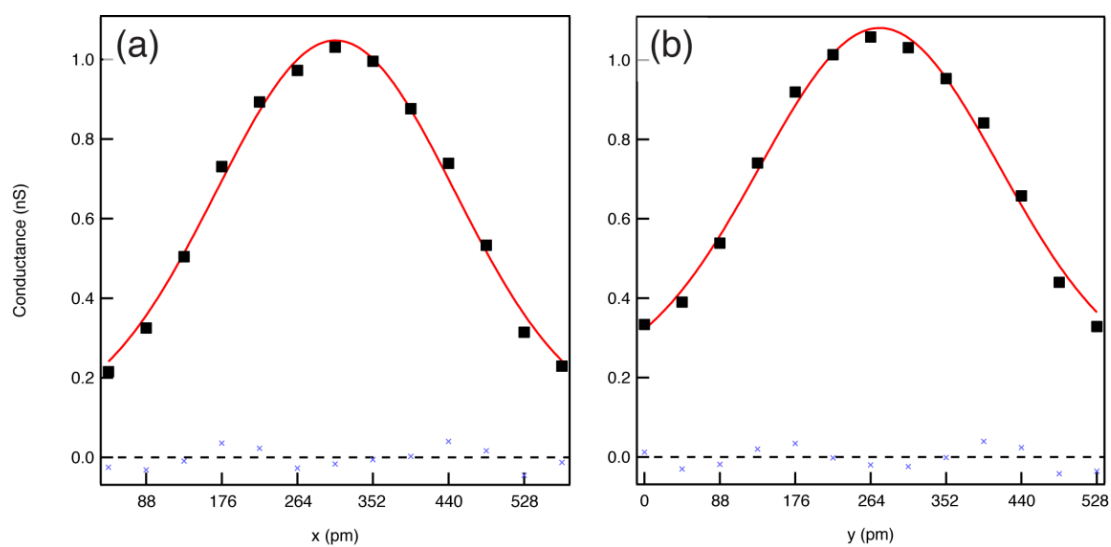

**Figure S1 (a) and (b) Horizontal (x) and vertical (y) line cuts, respectively, of the center Zn resonance in Figure 3j are shown as black squares. The red line is a 1D Gaussian fit to the data. We note the very good quality of the fit, which can be seen from the small difference between data and fit (blue crosses).**

Figure S2 and S3 show the same Zn resonances as Figure 3 in the main text. For each Zn resonance, a white dot has been placed at the center of the two-dimensional Gaussian fit, which we use as the Zn atom coordinate in our analysis.

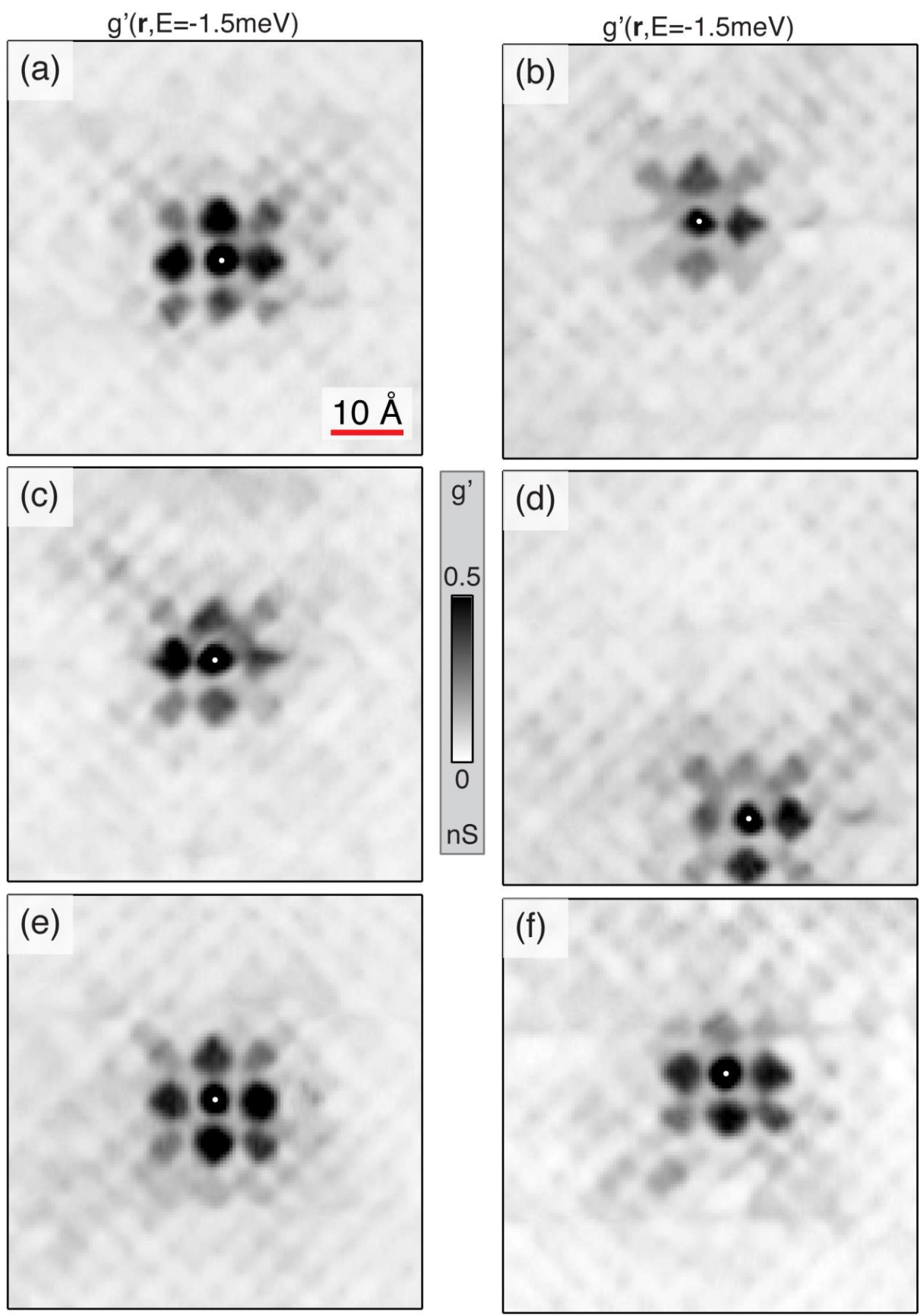

**Figure S2 (a)-(f)** Differential conductance data g'(r, E = -1.5meV) after distortion correction. The Zn resonances are the same as those in Figure 3 of the main text (3b through 3l). The coordinates of the Zn atoms as determined from 2D Gaussian fits are shown as white dots.

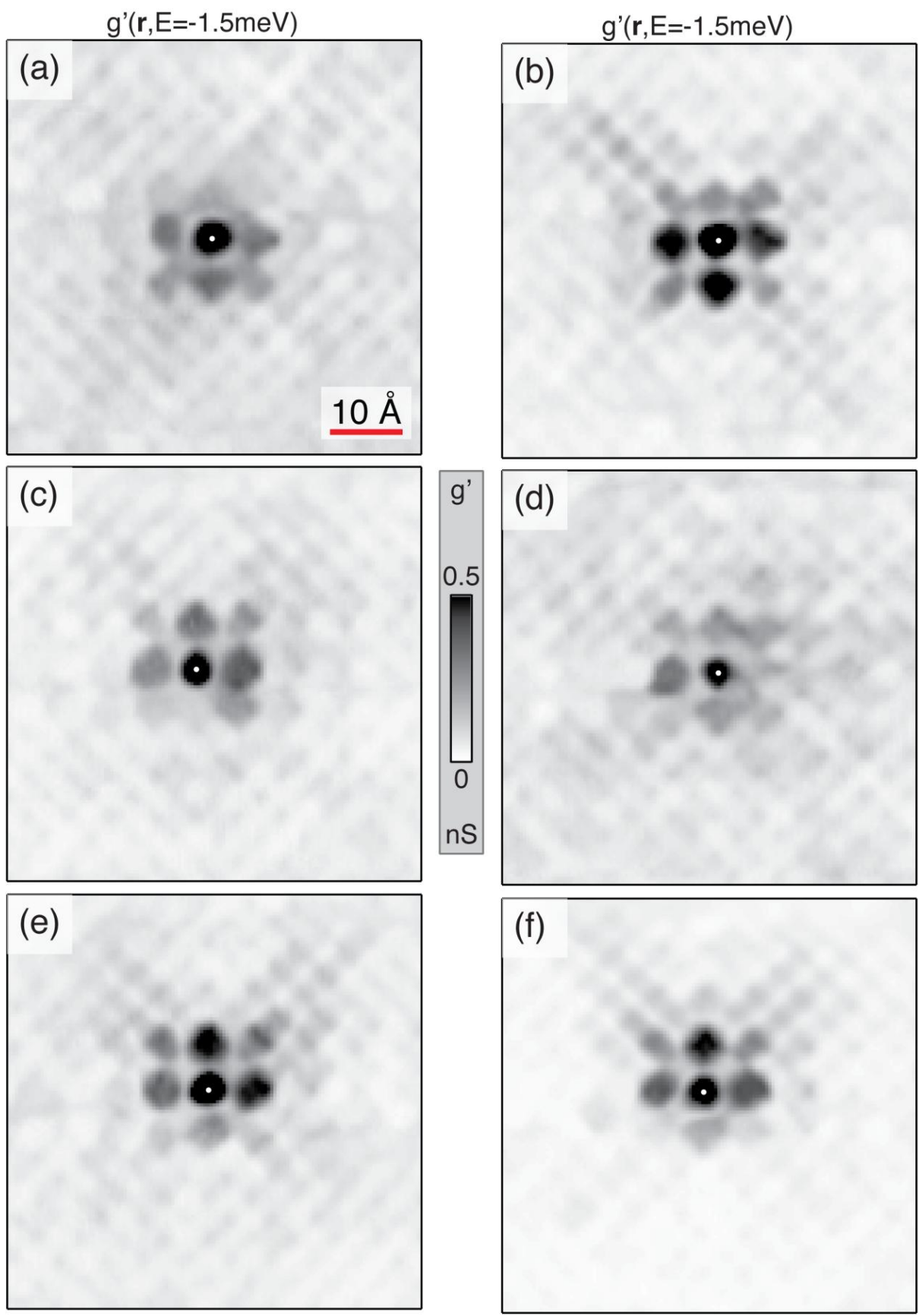

**Figure S3 (a)-(f)** Differential conductance data g'(r, E = -1.5meV) after distortion correction. The Zn resonances are the same as those in Figure 3 of the main text (3n through 3x). The coordinates of the Zn atoms as determined from 2D Gaussian fitting are shown as white dots.

## II. Dependence of results on scanning direction

It is important to understand the origin of the 2% ($a_0$) average displacement between Zn and Bi atoms in order to assess its physical consequences. To determine whether the origin of the shift is a property of the crystal lattice or due to measurement error, we study data from the same Zn atoms, now acquired with 180° rotated scan direction (RSD) to those previously reported in the main text. The effect of RSD on data acquisition is related to the change in some of the systematic drifts affecting tip travel times. Therefore, if a comparison of (Zn, Bi) displacements from RSD and original data results in equivalent displacement vectors for a given Zn resonance, the shift is likely originating from an actual displacement between the BiO and $CuO_2$ layers. In the opposite case, if, for a given Zn resonance, the RSD and original displacement vectors are not related, then the result is likely to be dominated by the apparatus or measurement limitations at these small length scales.

The RSD data were acquired in same FOV (see Figure 2) using the same acquisition parameters used for the data reported in the main text (except for the scanning direction itself). The RSD data were further processed and analyzed in a similar manner to the original data – in particular, distortion correction and 2D Gaussian fit to central Zn resonances. The results are shown in Figure S4. Contrary to the expectation for a real shift between the BiO and $CuO_2$ planes, we observe no obvious relation between the resulting (Zn, Bi) displacement vectors in the two measurements, in support of the scenario that the average shift is not a physical property of the lattice, but is rather caused by probe/measurement limitations at these picometer length scales.

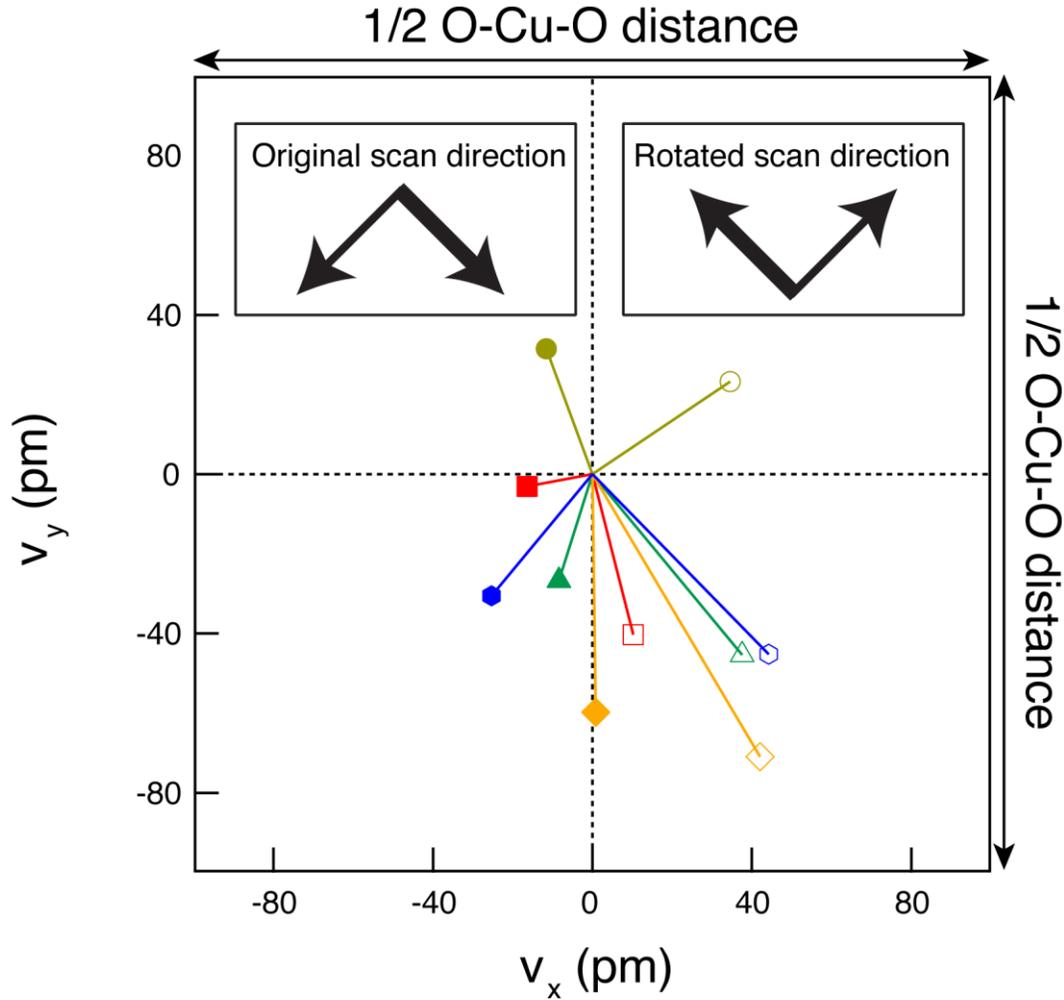

**Figure S4** Some of the (Zn, Bi) displacement vectors in Figure 4b from the manuscript (here as closed symbols) and the equivalent displacement vectors determined from a 180° RSD map in the same FOV (open symbols). Lines/symbols of similar color correspond to the same Zn and Bi atoms. We note that vectors (lines) of the same color have very different magnitude and orientation. Black arrows indicate fast (thicker) and slow (thinner) scan directions in original maps and their equivalent 180° RSD maps. Bi atoms are superimposed at the origin.

### III. Comments on the results by *Lawler et al.* [2]

If one of the atoms of some square lattice $L'(\mathbf{r})$ lies exactly at the origin of the coordinate system of the image, then $L'(\mathbf{r})$ can be treated symmetrically for the purposes fourier transform analysis. A symmetric $L'(\mathbf{r})$ results in a zero imaginary part of its Fourier transform, $\tilde{L}'(\mathbf{q})$, at the Bragg wavevectors, $\operatorname{Im} \tilde{L}'(\mathbf{Q}_{x,y}) = 0$, where $\mathbf{Q}_{x,y}$ are the Bragg wavevectors in the x- and y-directions, respectively. For an asymmetric image, one which does not have a lattice point at the origin, Im $\tilde{L}'(Q_{x,y})$ has a finite value and, depending on the direction and magnitude of the shift, the amplitude of the real part of the Bragg peaks in the two perpendicular directions, Re $\tilde{L}'(Q_{x,y})$, differ from one another. We note that, in terms of the Fourier transform phase, a spatial shift of the lattice by $n\%$ ($a_0$) corresponds to a phase shift of $n\%$ ($2\pi$) at the Bragg wavevectors.

A lattice of two dimensional Gaussian "atoms" was simulated to assess the impact of a 2% ($a_0$) displacement between the BiO and CuO$_2$ layers in the measure of "nematicity" discussed in [2]. As expected, a zero lattice shift yields that Im $\tilde{L}'(Q_{x,y}) = 0$. Additionally, Re $\tilde{L}'(Q_{x,y})$ is identical in the x- and y-directions. We consider the case of a 2% ($a_0$) displacement either along the x- or y-direction. In this case, the measure of "nematicity" defined in Eq. S1 results in a ~1% difference between the values of Re $\tilde{L}'(Q_{x,y})$.

$$O_N^Q \equiv \frac{\text{Re } \tilde{L}'(Q_y) - \text{Re } \tilde{L}'(Q_x)}{\text{Re } \tilde{L}'(Q_y)} \sim 1\% \tag{S1}$$

Compared to the > 30% effect discussed in [2], the influence of the 2% ($a_0$) shift is negligible. In fact, a very large displacement, $\gtrsim$15% ($a_0$), would be required for $O_N^Q$ (Eq. S1) to return values comparable to the results discussed in [2].

Regarding the measure for inversion symmetry discussed in the main text (see Eq. S2), the same 2% ($a_0$) lattice shift in the x- or y-direction produces Im $\tilde{L}'(Q_{x,y}) \sim 13\% \times$ Re $\tilde{L}'(Q_{x,y})$, a value that is of the same order of magnitude as the physical effects one expects to measure in $O_I^Q$ (Eq. S2).

$$O_I^Q \equiv \frac{\text{Im } \tilde{L}'(Q_{x,y})}{\text{Re } \tilde{L}'(Q_{x,y})} \sim 13\% \tag{S2}$$

In light of this result, we regard **q**-space inversion symmetry studies to be, at the moment, mostly likely unfeasible for SI-STM studies of the PG phase in Bi$_2$Sr$_2$CaCuO$_{8+\delta}$. In other words, for valid SI-STM **q**-space inversion symmetry studies, the measurement should produce a systematic shift ideally no larger than

1% ($a_0$), for which case $O_I^Q \lesssim 6\%$ (Eq. S2). Importantly, the standard deviation around this estimated shift value is also required to be no larger than 1% or 2%.